\RequirePackage{fix-cm}
\documentclass[smallextended]{svjour3}       
\smartqed  
\usepackage{graphicx}
\usepackage[utf8]{inputenc}
\usepackage{amsmath,amssymb}
\usepackage{tikz}
\usepackage[normalem]{ulem}
\usepackage{indentfirst}
\usepackage[T1]{fontenc}
\usepackage{graphics}
\usepackage{xcolor}
\usepackage{epigraph, varwidth}
\usepackage{pdfpages}
\usepackage{setspace}
\usepackage[font=small,labelfont=bf]{caption}
\usepackage{float}
\usepackage{pgfplots}
\usepackage{epstopdf}
\usepackage{subfig}
\usepackage{afterpage}
\usepackage{array}
\usepackage{emptypage}
\usepackage[retainorgcmds]{IEEEtrantools}
\usepackage[acronym,nonumberlist,sort=use]{glossaries}
\usepackage{enumitem}
\usepackage[colorlinks=true, allcolors=black, citecolor=blue, urlcolor=blue]{hyperref}
\usepackage[super]{nth}
\usepackage{natbib}
\setcitestyle{aysep={}} 
\newcommand{\be}{\begin{equation}}
\newcommand{\ee}{\end{equation}}
\newcommand{\bal}{\begin{aligned}}
\newcommand{\eal}{\end{aligned}}
\newcommand{\beq}{\begin{eqnarray}}
\newcommand{\eeq}{\end{eqnarray}}
\newcommand{\ba}{\begin{array}}
\newcommand{\ea}{\end{array}}
\newcommand{\bs}{\boldsymbol}
\newcommand{\bi}{\begin{itemize}}
\newcommand{\ei}{\end{itemize}}

 \journalname{Celestial Mechanics and Dynamical Astronomy}
\AtBeginDocument{%
  \paperwidth=\dimexpr
    1in + \oddsidemargin
    + \textwidth
    + 1in + \oddsidemargin
  \relax
  \paperheight=\dimexpr
    1in + \topmargin
    + \headheight + \headsep
    + \textheight
    + 1in + \topmargin
  \relax
  \usepackage[pass]{geometry}\relax
}

\begin{document}

\title{Order-chaos-order and invariant manifolds in the bounded planar Earth-Moon system
}
%


\titlerunning{Chaos and manifolds in the Earth-Moon system}        

\author{Vitor M. de Oliveira \and Priscilla A. Sousa-Silva \and Iberê L. Caldas 
}


\institute{V. M. de Oliveira  \and I. L. Caldas \at
              USP - University of São Paulo, Institute of Physics\\
              Rua do Matão, 1371 - Edif. Basílio Jafet - CEP 05508-090 \\ 
              Cidade Universitária - São Paulo/SP - Brasil \\
              \email{vitormo@if.usp.br}           
              \and
           Priscilla A. Sousa-Silva \at
           UNESP - São Paulo State University, São João da Boa Vista \\
           Avenida Professora Isette Corrêa Fontão, 505 - CEP 13876-750\\
           Jardim das Flores - São João da Boa Vista/SP - Brasil
}

\date{Received: date / Accepted: date}

\maketitle

\begin{abstract}
In this work, we investigate the Earth-Moon system, as modeled by the planar circular restricted three-body problem, and relate its dynamical properties to the underlying structure associated with specific invariant manifolds.
We consider a range of Jacobi constant values for which the neck around the Lagrangian point $L_1$ is always open but the orbits are bounded due to Hill stability. First, we show that the system displays three different dynamical scenarios in the neighborhood of the Moon: two mixed ones, with regular and chaotic orbits, and an almost entirely chaotic one in between. We then analyze the transitions between these scenarios using the Monodromy matrix theory and determine that they are given by two specific types of bifurcations.
After that, we illustrate how the phase space configurations, particularly the shapes of stability regions and stickiness, are intrinsically related to the hyperbolic invariant manifolds of the Lyapunov orbits around $L_1$ and also to the ones of some particular unstable periodic orbits.
Lastly, we define transit time in a manner that is useful to depict dynamical trapping and show that the traced geometrical structures are also connected to the transport properties of the system.

\end{abstract}

\section{Introduction}
\label{sec:intro}

The dynamics of nonintegrable Hamiltonian systems is characterized by the coexistence of both chaotic and regular motion or by the complete lack of the latter. The first case corresponds to a \emph{mixed phase space}, which is composed of regions of stability and areas filled by chaotic orbits, while the second one corresponds to a large chaotic sea in the phase space \citep{Lichtenberg1992}.

A good understanding of this type of system comes from analyzing how such phase space scenarios are affected by the value of the constants of motion.
A complete description, however, also involves the underlying geometrical structures, which are related to the hyperbolic invariant manifolds associated with unstable periodic orbits in the chaotic sea and whose properties can influence the systems' dynamics.

Chaotic behavior is a common feature in Celestial Mechanics since many systems are represented by a nonintegrable Hamiltonian function. Such behavior is associated with, for example, the motion of asteroids and of the solar system itself \citep{Poincare1890,Laskar1989,Ferraz1999}. Invariant manifolds have been extensively investigated in this field as well and employed in a variety of applications, ranging from natural transport to space mission design \citep[e.g.,][]{Koon2008,Gawlik2009,Perozzi2010}.

In this work, we adopt the planar Circular Restricted Three-Body Problem (CRTBP) as a model to investigate the dynamical properties of the Earth-Moon system.
Specifically, we are interested in how dynamical objects, such as periodic orbits, invariant tori and hyperbolic invariant manifolds, behave as the value of the Jacobi constant varies.
Our analysis is focused on a Poincaré section in the neighborhood of the Moon.

Periodic solutions of the CRTBP were widely studied considering the mass parameter that corresponds to the Earth-Moon system \citep[e.g.,][]{Szebehely1967,Broucke1968,Henon1997}. We are especially interested in the \emph{Lyapunov orbits} around $L_1$ and the \emph{Low Prograde Orbits} and \emph{Distant Prograde Orbits}, both of which are part of the \emph{direct periodic orbits}
around the Moon \citep{Restrepo2018}.
Given that connections between these and other periodic solutions of the system provide low-cost transfers between different regions of the phase space \citep{Mingotti2012,Cox2020}, it is useful to know which orbits coexist for the same Jacobi constant value in order to define which transfers are accessible \citep{Folta2015}.

It is worth mentioning that the periodic and transfer orbits of the planar CRTBP are also used as a reference for calculating orbits in more complicated models which considers, for instance, the mass of the third particle, the eccentricity of the lunar orbit and the influence of the Sun \citep{Szebehely1967, Leiva2008}.

In some situations, invariant manifolds are also responsible for \emph{stickiness}, where chaotic orbits in Hamiltonian systems spend a significant amount of time around a regular region  \citep{Contopoulos2004}. The occurrence of stickiness in the system then implies a higher concentration of orbits behaving similarly around a given area of the phase space. This phenomenon is associated with invariant manifolds, e.g., in the dynamics of spiral galaxies \citep{Contopoulos2010}.

In the present manuscript, we consider a range of values for the constant of motion in which all orbits analyzed are bounded within the system and we use numerical tools to obtain the periodic orbits and their respective invariant manifolds.
In order to investigate stickiness, we are also interested in the unstable periodic orbits that are formed from the destruction of an invariant KAM torus and that live in the neighborhood of a given regular region, additionally to the periodic orbits mentioned before.

We show how the invariant manifolds of the main periodic orbits of the system occupy the available area of the phase space as the Jacobi constant changes, a relevant aspect for practical purposes. Furthermore, we present a visual description of how these structures affect the transport of the system and, consequently, the chaotic dynamics, a relevant aspect for natural phenomena. In summary, our results illustrate how geometrical structures relate to the phase space scenarios, thus contributing to understanding the fundamental connection between dynamics and geometry in the Earth-Moon system.

This paper is organized as follows. In Sec. \ref{sec:system} we present the planar CRTBP and its dynamical features. In Sec. \ref{sec:chaos} we describe the phase space configuration for the considered range of Jacobi constant and discuss the bifurcations that occur in the stability regions. In Sec. \ref{sec:manifolds} we trace the invariant manifolds associated with the Lyapunov orbits around $L_1$ and illustrate their relation to the phase space configuration. Later, we consider the stickiness effect by tracing the manifolds associated with selected unstable periodic orbits in the mixed phase space scenarios. In Sec. \ref{sec:transport} we define transit time in a suitable manner and examine the transport properties of the system. Finally, we give our conclusions in Sec. \ref{sec:conclusions}.

\section{Physical system}
\label{sec:system}

The framework we use is the planar CRTBP, which provides a good first approximation to the dynamical behavior of the Earth-Moon system \citep{Murray1999}. It describes the motion of a body of negligible mass under the gravitational influence of two massive bodies moving in circular orbits around their common center of mass. 

We assume that the third body moves in the same plane as the two-body system.
This is a useful assumption because both the planar version of the problem and the geometrical structures that we deal with in this work have the advantage of being naturally represented in a two-dimensional surface of section. 

In a synodic reference frame, which rotates with the same frequency as the system formed by the primaries and is centered at its center of mass, the dimensionless equations of motion in terms of the variables $(x,y,\dot{x},\dot{y},t)$ are\footnote{For the equations of motion in Hamiltonian form see, e.g., \cite{Belbruno2004}.}

\be
\begin{aligned}
\ddot{x}-2\dot{y} &= \dfrac{\partial{\Omega}}{\partial x},\\
\ddot{y}+2\dot{x} &= \dfrac{\partial{\Omega}}{\partial y},\\
\end{aligned}
\label{eq:motion_rotational}
\ee

\noindent where the pseudo-potential $\Omega$ is given by

\be
\Omega = \frac{1}{2}(x^2+y^2)+\frac{1-\mu}{\sqrt{(x+\mu)^2+y^2}}+\frac{\mu}{\sqrt{(x-(1-\mu))^2+y^2}}.
\label{eq:omega}
\ee

\begin{figure*}
\centering
\includegraphics[scale=0.7]{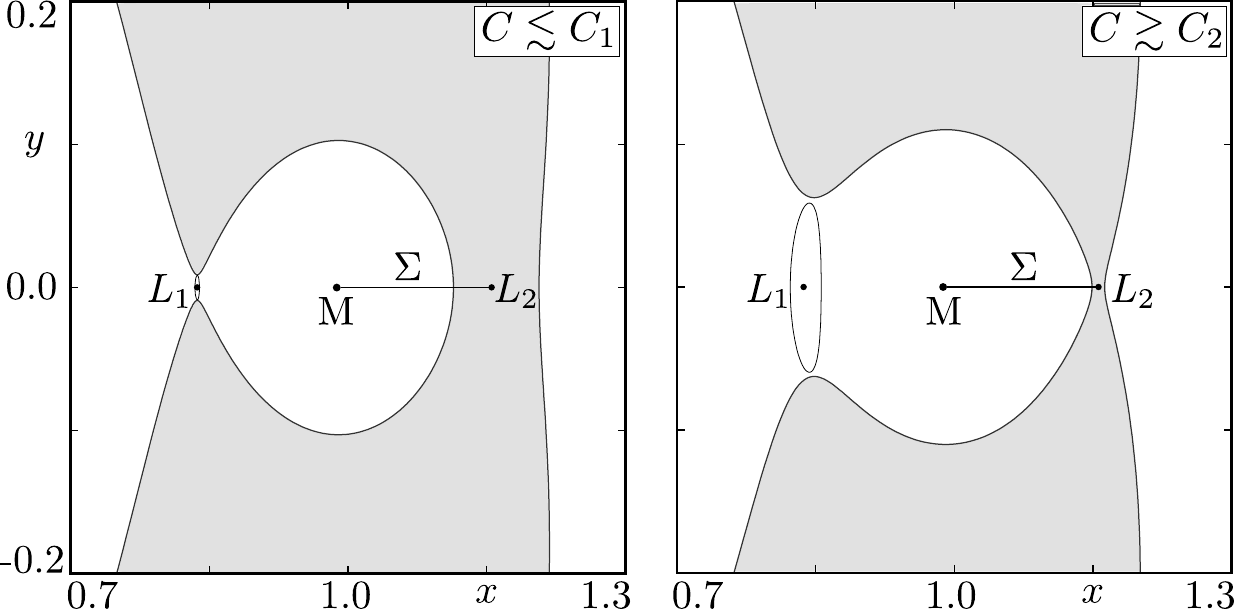}
\caption{Physical model in the vicinity of the Moon. The third body moves in the white area, the Hill region, and $\Sigma$ represents the Poincaré surface where the orbits are analyzed. As the Jacobi constant $C$ goes from $C_1$ to $C_2$, the neck around the Lagrangian point $L_1$ becomes larger. The Lyapunov orbit is also depicted for both situations.}
\label{fig:hill}
\end{figure*}

The primaries are located at $(-\mu,0)$ and $(1-\mu,0)$, with $\mu$ being the \emph{mass parameter}, the ratio between the mass of the least massive primary and the system's total mass. For the Earth-Moon system, we have $\mu=1.2150\times10^{-2}$. A schematic representation of the system around the Moon is presented in Fig. \ref{fig:hill}, along with some important concepts which are addressed later in this paper.

There is one unstable Lagrangian equilibrium point on each side of the Moon, namely $L_1$ (left) and $L_2$ (right). These equilibrium points are collinear to the primaries and their positions in the $x$-axis depend only on $\mu$ \citep{Gomez2001}. For the Earth-Moon system, their locations are given by $x_{L_1}\approx0.8369$ and $x_{L_2}\approx1.1556$.

The system has one constant of motion, called \emph{Jacobi constant} $C$, which is given by

\be
C = 2\Omega-\dot{x}^2-\dot{y}^2,
\label{eq:constant}
\ee

\noindent and, as the constant $C$ is a constraint of the system, the dynamics effectively occurs in a three-dimensional subspace. Additionally, since $\dot{x}^2+\dot{y}^2>0$, Eq.~\eqref{eq:constant} defines the area accessible to the third body in the coordinate space $x$-$y$ for a given C, 

\be
\mathcal{H}=\{(x,y)\in\mathbb{R}^2~|~2\Omega-C>0\},
\label{eq:hill_region}
\ee

\noindent called the \emph{Hill region}. In Fig. \ref{fig:hill}, $\mathcal{H}$ is represented by the white area. Furthermore, $C_1 \approx 3.1883$ and $C_2 \approx 3.1722$ are the Jacobi constants at the Lagrangian points $L_1$ and $L_2$, respectively.

\begin{figure*}
\centering
\includegraphics[scale=0.7]{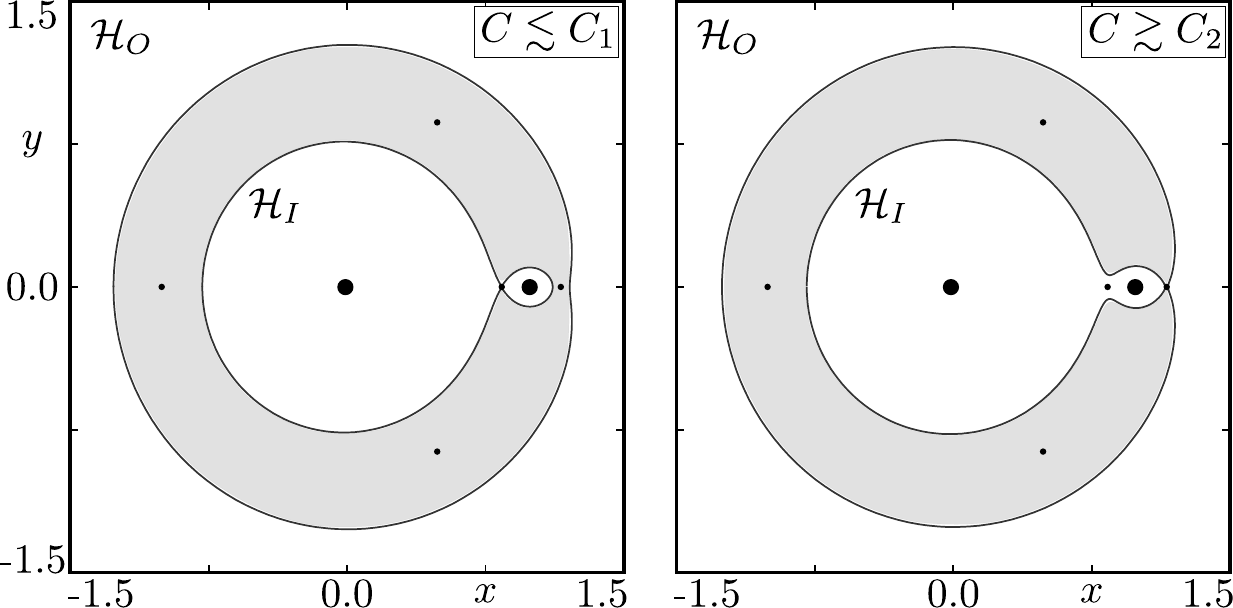}
\caption{Accessible area for the full system. The Hill region $\mathcal{H}$ is composed by an inner $\mathcal{H}_I$ and an outer $\mathcal{H}_O$ area, which are disconnected. The five Lagrangian points are represented by the small black circles, while the primaries are represented by the bigger ones. For $C_1>C>C_2$, $L_1$ is the only Lagrangian point in $\mathcal{H}$.}
\label{fig:hill_full}
\end{figure*}

It is important to note that, for the range of Jacobi constant chosen in this work, $C_1 > C > C_2$, $\mathcal{H}$ is divided in two disconnected areas, that we define as the inner region $\mathcal{H}_I$ and the outer region $\mathcal{H}_O$, as shown in Fig.~\ref{fig:hill_full}. Consequently, the orbits that lie in the vicinity of either primary are bounded and cannot exit the system (\emph{Hill stability}). 

The boundaries of both $\mathcal{H}_I$ and $\mathcal{H}_O$ are given by the \emph{zero velocity curves}, which can be obtained by setting $\dot{x}=\dot{y}=0$ in Eq.~\eqref{eq:constant}. Furthermore, the Lagrangian point $L_1$ separates the inner Hill region $\mathcal{H}_I$ into two realms,

\be
\begin{aligned}
\text{\emph{Earth's realm}} &= \{(x,y)\in\mathcal{H}_I~|~x<x_{L_1}\},\\
\text{\emph{Moon's realm}} &= \{(x,y)\in\mathcal{H}_I~|~x>x_{L_1}\},
\end{aligned}
\ee

\noindent where $x_{L_1}$ is the position of $L_1$ in the $x$-axis.

Since our analyses involve numerical calculations, it is necessary to deal with the singularities in Eq.~\eqref{eq:omega}. This is achieved by using the \emph{Levi-Civita transformation} \citep{Szebehely1967}. Let $(u,v,u^{ \prime},v^{\prime},\tau)$ be the new set of variables in the system and let us define $z = x + i y$ and $\omega = u + i v$. The transformations are then given by $z = \omega^2 -\mu + 1$ for regularization in a vicinity of the Moon and $z = \omega^2 -\mu$ for regularization in a vicinity of the Earth. In both cases, the relation between the time variables is given by $dt=4(u^2+ v^2)d\tau$.

In the new set of variables, the equations of motion, Eq.~\eqref{eq:motion_rotational}, become

\be
\begin{aligned}
u^{\prime\prime}-8(u^2+v^2)v^{\prime} &= \dfrac{\partial V}{\partial u},\\
v^{\prime\prime}+8(u^2+v^2)u^{\prime} &= \dfrac{\partial V}{\partial v},\\
\end{aligned}
\label{eq:motion_regularized}
\ee

\noindent where the new pseudo-potential $V$ is 

\begingroup
\setlength\abovedisplayskip{0pt}
\begin{multline}
V_{M}(u,v)=4\mu+2(u^2+v^2)\Bigg\{{(u^2+v^2)}^2+2(1-\mu)(u^2-v^2)
\\
\left.+(1-\mu-C)+\dfrac{2(1-\mu)}{\sqrt{1+{(u^2+v^2)}^2+2(u^2-v^2)}}\right\}
\label{eq:V_Moon}
\end{multline}
\endgroup

\noindent for the Moon and,
 
\begingroup
\setlength\abovedisplayskip{0pt}
\begin{multline}
V_{E}(u,v)=4(1-\mu)+2(u^2+v^2)\Bigg\{{(u^2+v^2)}^2-2\mu(u^2-v^2)
\\
\left.+(\mu-C)+\dfrac{2\mu}{\sqrt{1+{(u^2+v^2)}^2-2(u^2-v^2)}}\right\}
\label{eq:V_Earth}
\end{multline}
\endgroup

\noindent for the Earth.

The regularization procedure is performed locally about the singularities. In practice, we establish two radii with values $\delta_{M}=1.00\times10^{-2}$  around the Moon and $\delta_{E}=3.67\times10^{-2}$ around the Earth. We then switch between equations \eqref{eq:motion_rotational} and \eqref{eq:motion_regularized} as soon as the orbits are detected to enter or exit one of these regions. Since the integration steps are kept small and $\delta_{E}$ and $\delta_{M}$ are large enough, it is not necessary to compute the exact point of intersection between the orbits and the circle defined by each regularization radius. In physical units, we have $\delta_{M} = 3844.0$ km and $\delta_{E}=14107.5$ km, while the mean radii of the primaries are $r_{M}=1737.4$ km and $r_{E}=6371.0$ km.\footnote{Values from \url{https://nssdc.gsfc.nasa.gov/planetary/factsheet/moonfact.html}.}

In both transformations, the system is re-centered to one of the primaries and we can verify in equations \eqref{eq:V_Moon} and \eqref{eq:V_Earth} that $V$ is finite when $(u,v)\to(0,0)$, thus removing the singularities from these locations.

\section{Order-chaos-order}
\label{sec:chaos}

We now proceed to study the dynamical properties in the vicinity of the Moon. In order to do so we choose a surface of section $\Sigma$ between the Moon and $L_2$ defined by

\be
\Sigma=\{\bs{x} = (x,y,\dot{x},\dot{y})~|~1-\mu< x<x_{L_2},~y=0,~\dot{y}>0\},
\ee

\noindent where $x_{L_2}$ is the position of $L_2$ in the $x$-axis. In Fig. \ref{fig:hill}, we depict $\Sigma$ for $C\lesssim C_1$ and $C\gtrsim C_2$.

\begin{figure*}
\centering
\includegraphics[scale=0.75]{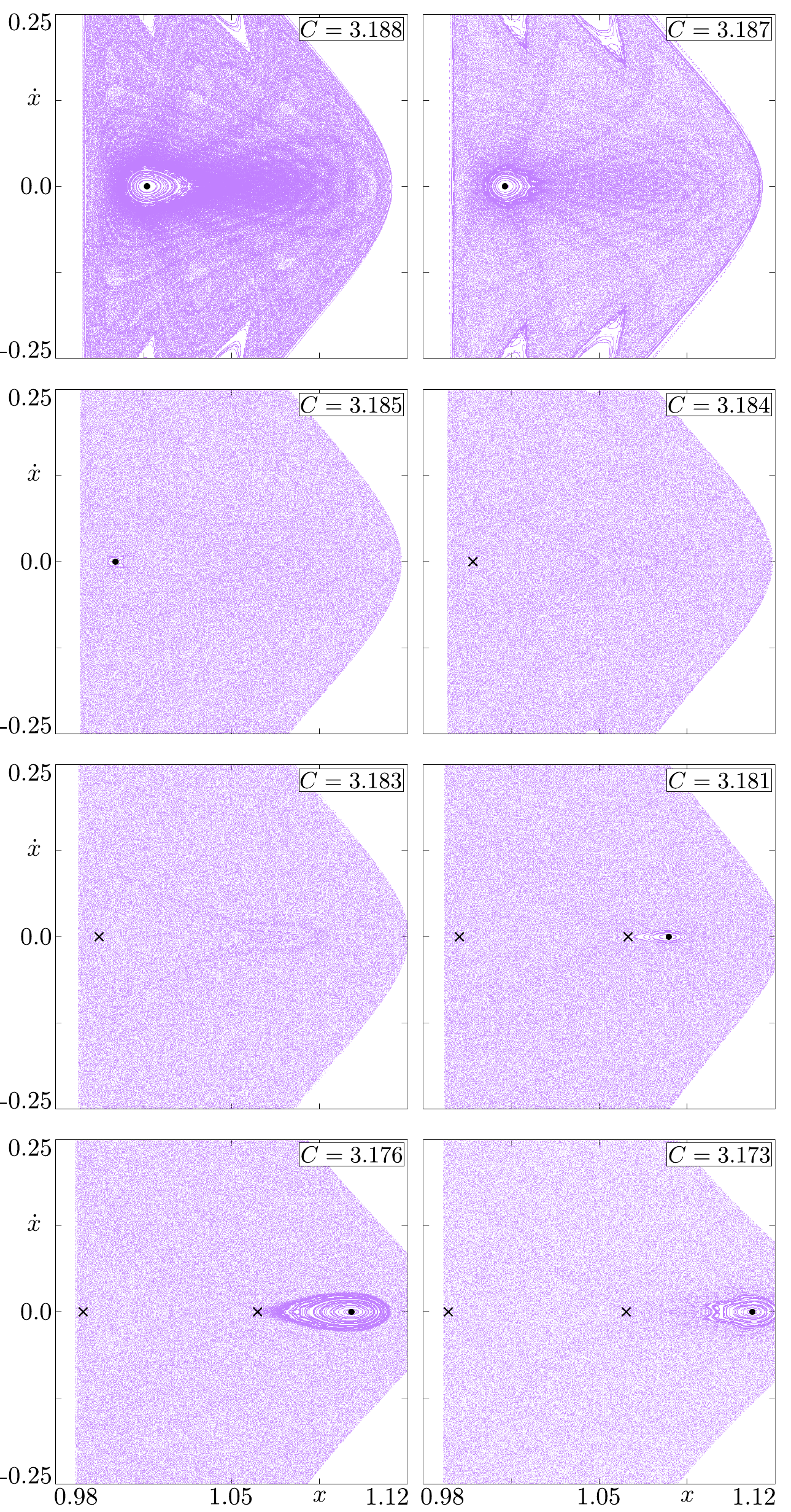}
\caption{Phase space in the surface of section $\Sigma$ for the selected range of Jacobi constant $C$. The system goes from and back to a mixed scenario but with different stickiness behavior. The black circles mark the location of stable periodic orbits, while the black crosses indicate the location of unstable periodic orbits.}
\label{fig:phase_space}
\end{figure*}

Fig. \ref{fig:phase_space} shows the system's phase space $x$-$\dot{x}$ for different values of $C$. The initial conditions are chosen in a $36$ by $36$ grid on $\Sigma$ and the orbits are integrated up to $t=5\times10^3$ both forward and backward in time, which corresponds to approximately 748.5 years. Numerical integration of the equations of motion are carried out using the explicit embedded Runge-Kutta Prince-Dormand 8(9) \citep{Galassi2001} and errors associated with the Jacobi constant along the orbit and with the intersection between orbit and surface of section are kept below $10^{-10}$.

The first feature we observe is the existence of three different scenarios as the Jacobi constant is decreased: I. ($C=3.188$, $3.187$ and $3.185$) the system presents a mixed phase space and the region of stability decreases in size; II. ($C=3.184$ and $3.183$) all orbits analyzed are chaotic and hence the former stability region was destroyed; III. ($C=3.181$, $3.176$ and $3.173$) the phase space becomes mixed again with the creation, enlargement and subsequent slight decrease in size of a new stability region.

We can use Newton's Method and the symmetry of the model to calculate both stable and unstable periodic orbits in the system for adequate initial conditions. In order to understand then what happens with the stability regions in both mixed phase space scenarios, we follow the periodic orbits in each case and study their stability by computing the eigenvalues of their respective Monodromy matrices.

The Monodromy matrix has four eigenvalues, two of which are always unitary. The remaining two eigenvalues determine the stability of the periodic orbit as follows: if the orbit is stable, the eigenvalues are complex conjugate to each other; however, if the orbit is unstable, the eigenvalues are real and one is the inverse of the other \citep{Meyer2008}.

\begin{figure*}
\centering
\subfloat[Scenario I]{\includegraphics[scale=0.70]{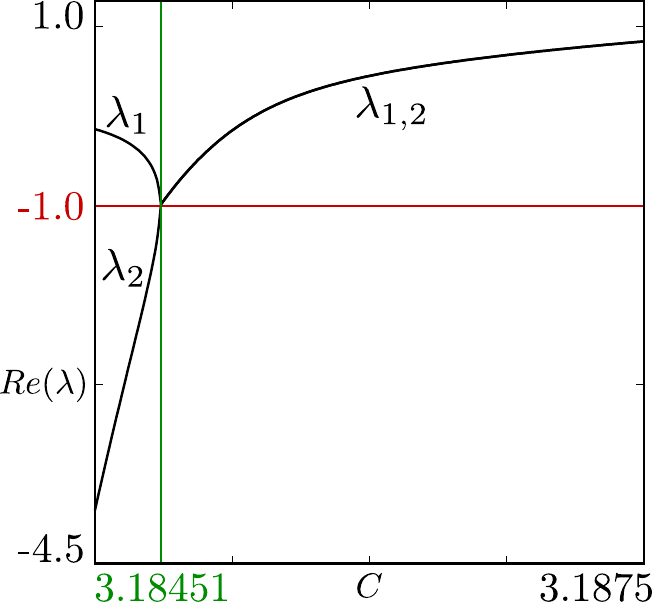}\label{subfig:inner_eigenvalues}}~~
\subfloat[Scenario III]{\includegraphics[scale=0.70]{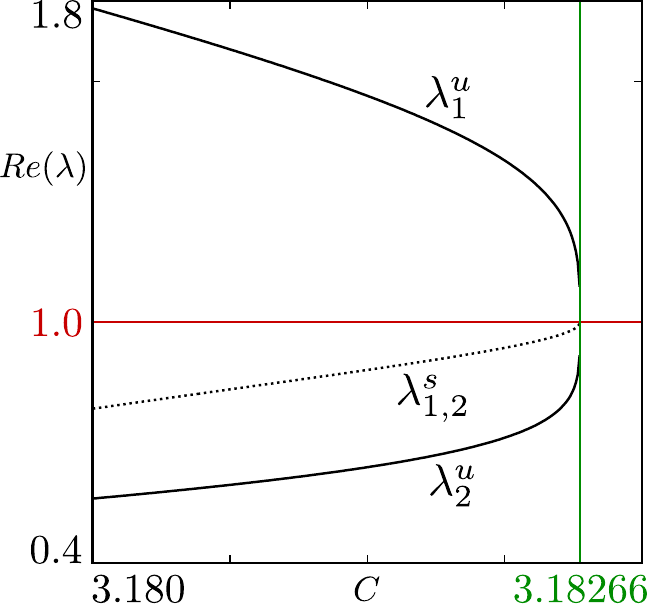}\label{subfig:outer_eigenvalues}}\\
\subfloat[Scenario III]{\includegraphics[scale=0.70]{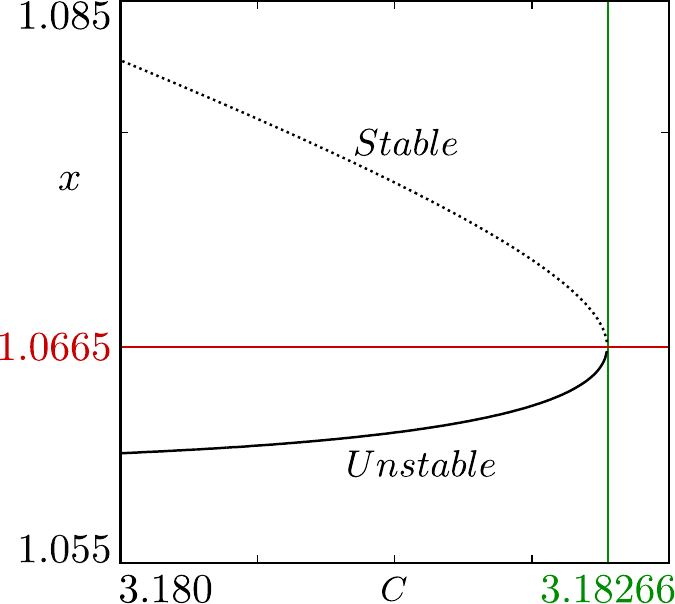}\label{subfig:outer_position}}
\caption{Bifurcation analysis for both mixed phase space scenarios. The real part of the eigenvalues $Re(\lambda)$ as a function of the Jacobi constant $C$ are shown for \protect\subref{subfig:inner_eigenvalues} the center orbit in Scenario I and \protect\subref{subfig:outer_eigenvalues} the period-1 stable and unstable orbits in Scenario III. In \protect\subref{subfig:outer_position}, the orbits in Scenario III are shown to collide as their $x$-axis components tend to the same value.}
\label{fig:bifurcation}
\end{figure*}

Since we are investigating the dynamics of the system on $\Sigma$, we consider the period of a periodic orbit as the number of times said orbit crosses our surface of section before closing in on itself.
In Scenario I, there is one periodic orbit of period 1 which is initially located at the center of the stability region (see Fig.~\ref{fig:phase_space}). In Fig. \ref{subfig:inner_eigenvalues} we evaluate the real part of both eigenvalues of this orbit which are associated with stability as a function of the Jacobi constant. We observe that the orbit is stable for $C=3.1875$ and it eventually becomes unstable as $C$ is lowered. We have, in this case, a direct or inverse bifurcation\footnote{The direction of the bifurcation determines the stability of a new periodic orbit which appears outside $\Sigma$ and hence it is not relevant to our analysis.} \citep{Contopoulos2004}, which happens at approximately $C^1_ {bif}=3.18451$.

In Scenario III, there are two more periodic orbits of period 1: the stable one at the center of the stability region and its unstable counterpart to the left of it, just outside the stability region and inside the chaotic sea (see Fig.~\ref{fig:phase_space}). We perform the same analysis as before for both orbits, but this time we increase the Jacobi constant. The results are shown in Fig. \ref{subfig:outer_eigenvalues}. For $C=3.180$, all four eigenvalues are distinct and, as $C$ is increased, they all tend to the same value. We have, in this case, a saddle-node bifurcation \citep{Contopoulos2004}, which happens at approximately $C^2_{bif}=3.18266$. After the bifurcation is reached, both periodic orbits disappear. In Fig. \ref{subfig:outer_position} we present the position in the $x$-axis of both orbits up until their collision.

We note from Fig. \ref{fig:bifurcation} that the eigenvalues go through $-1$ in Scenario I and to $1$ in Scenario III. Hence, the trace of the Monodromy matrix goes to $0$ and $4$, respectively, both of which indicate the occurrence of a bifurcation in two-degree of freedom Hamiltonian systems \citep{Aguiar1987}.

The families of periodic orbits that are presented in Fig.~\ref{fig:phase_space} and analyzed in Fig.~\ref{fig:bifurcation} belong to a class of direct periodic orbits around the smaller primary, the Moon, which is referred to as the \emph{g class} \citep{Szebehely1967}.
The family in Scenario I along with the stable family in Scenario III are formed by the \emph{Low Prograde Orbits}, while the unstable periodic orbits in Scenario III are the \emph{Distant Prograde Orbits} \citep{Restrepo2018}.

The second feature which stands out in Fig. \ref{fig:phase_space} is the difference in the stickiness behavior in both mixed phase space scenarios.
In Scenario III, there is a higher orbit concentration just about the stability region as is usually the case. However, for higher values of $C$ in Scenario I, the stickiness effect reaches deep into the chaotic sea and far from the stable portion of phase space, which suggests that it is being caused by invariant manifolds associated with unstable periodic orbits around the stability region \citep{Contopoulos2010}.

We present a summary of the three dynamical scenarios in Tab. \ref{tab:scenarios}. The type \emph{order} indicates the presence of stability regions in the system.
As discussed before, the Hill region $\mathcal{H}$ is composed of two disconnected areas and it is important to note here that it remains as such in all scenarios. 

\renewcommand{\arraystretch}{1.2}
\begin{table}[h]
\caption{Overview of the three different scenarios that are present in the system.}
\label{tab:scenarios}  
\centering
\begin{tabular}{cccc}
\hline\noalign{\smallskip}
Scenario & Range & Type & Stickiness  \\
\noalign{\smallskip}\hline\noalign{\smallskip}
I & $~~C_1>C>C_{bif}^1$ & order & non-localized \\
II & $C_{bif}^1>C>C_{bif}^2$ & chaos & absent \\
III & $C_{bif}^2>C>C_2~~$ & order & localized \\
\noalign{\smallskip}\hline
\end{tabular}
\end{table}

\section{Invariant manifolds}
\label{sec:manifolds}

The Lagrangian point $L_1$ is the only equilibrium of the system which is inside the Hill region for the range of Jacobi constant that we considered. Furthermore, there exists an uniparametric family of unstable periodic orbits around this point, namely the \emph{Lyapunov orbits}. We are able to calculate a Lyapunov orbit for any value of $C$ using a continuation method along with the linear solution around $L_1$ \citep{Gomez2001}. For illustration, the orbits corresponding to $C = 3.1880\lesssim C_1$ and $C = 3.1725\gtrsim C_2$ are shown in Fig. \ref{fig:hill}.

Let $\bs{p}$ be a point of the unstable periodic orbit $\alpha$. As described in Sec. \ref{sec:chaos}, the Monodromy matrix calculated at $\bs{p}$ has a pair of real eigenvalues which determine the orbit's stability.
These eigenvalues, with moduli lower and greater than one, are related to eigenvectors that define a stable and an unstable direction, respectively. Therefore, there is a set of orbits that originate in a neighborhood of $\bs{p}$ and that tend to it as time goes to $\pm\infty$. If we extend this set to the whole space, we define the \emph{stable manifold} $W^s(\bs{p})$ and the \emph{unstable manifold} $W^u(\bs{p})$ associated with $\bs{p}$. Formally, we write

\be
\begin{aligned}
W^s(\bs{p})&=\{\bs{x}\in U\subset \mathbb{R}^4 \ | \ \varphi_ t(\bs{x})\to\bs{p} \ \text{as} \ t\to\infty\},\\
W^u(\bs{p})&=\{\bs{x}\in U\subset \mathbb{R}^4 \ | \ \varphi_ {t}(\bs{x})\to\bs{p} \ \text{as} \ t\to-\infty\},\\
\end{aligned}
\label{eq:manifold_1d_definition}
\ee

\noindent where $\varphi_t(\bs{x})$ is the solution of the system at time $t$ with initial condition $\bs{x}$. We can then define the stable manifold $W^s(\alpha)$ and unstable manifold $W^u(\alpha)$ associated with the unstable periodic orbit $\alpha$ as

\be
\begin{aligned}
W^{s,u}(\alpha)=\bigcup_{\bs{p}\in\alpha}W^{s,u}(\bs{p}).\\
\end{aligned}
\label{eq:manifold_2d_definition}
\ee

To numerically trace $W(\alpha)$, we first calculate one Monodromy matrix eigenvector and then propagate it to the other points of the discretized orbit $\alpha$ by multiplying it with the Transition matrix. We then take one initial condition on each vector with a distance of $10^{-6}$ from the orbit and integrate them forward or backward in time, depending on the eigenvector stability. Both $W^s(\alpha)$ and $W^u(\alpha)$ have two branches that are associated to an eigenvector and to its counterpart in the opposite direction. 

\begin{figure*}
\centering
\subfloat[]{\includegraphics[scale=0.7]{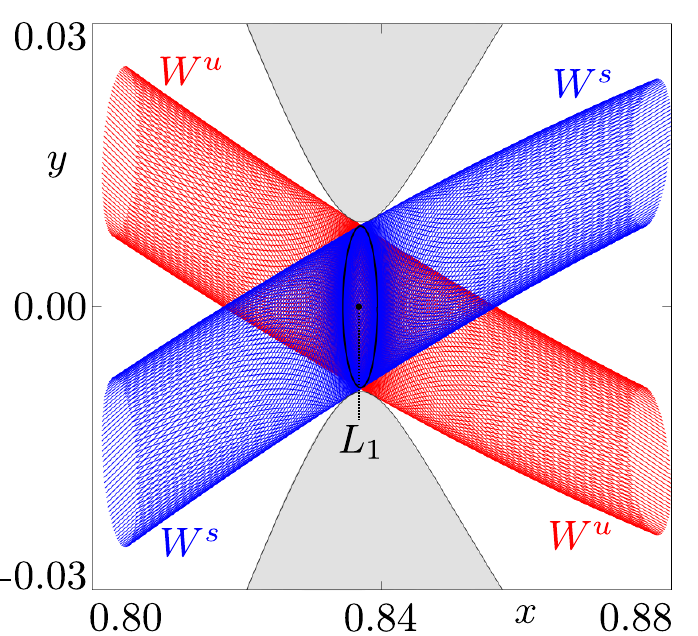}
\label{subfig:3_188_zoom}} \ \ \ \ \ 
\subfloat[]{\includegraphics[scale=0.7]{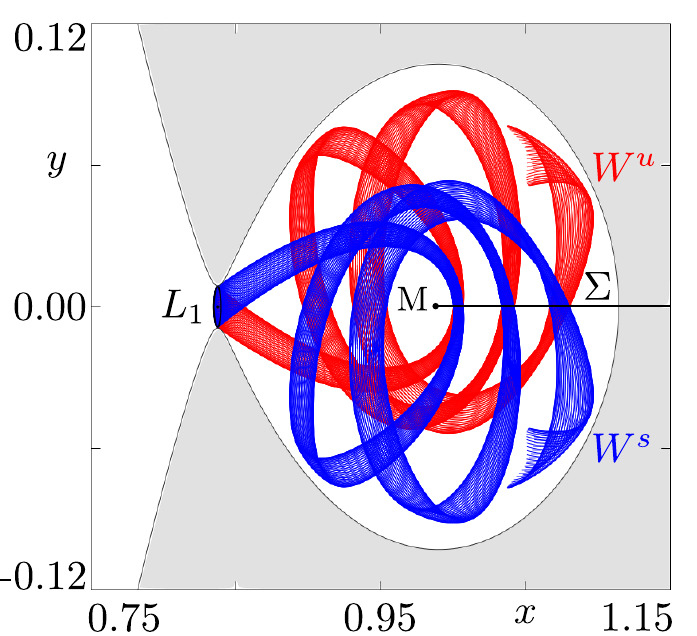}
\label{subfig:3_188_full}}
\caption{Projection on the coordinate space $x$-$y$ of the invariant manifolds $W(L)$ associated with the Lyapunov orbit for $C=3.188$. \protect\subref{subfig:3_188_zoom} Close-up near the Lyapunov orbit showing both branches with cylindrical shapes. \protect\subref{subfig:3_188_full} Evolution of the right branches crossing the surface of section $\Sigma$ inside the lunar realm.}
\label{fig:manifold_drawing_new}
\end{figure*}

Due to the fact that the dynamics in our system effectively occurs in a three-dimensional subspace, $\alpha$ is an one-dimensional curve and $W(\alpha)$ are two-dimensional surfaces that are locally homeomorphic to cylinders \citep{Ozorio1990}. In Fig.~\ref{fig:manifold_drawing_new}, we present the invariant manifolds $W(L)$ associated with the Lyapunov orbit for $C=3.188$ projected onto the coordinate space $x$-$y$.

The manifolds in Fig.~\ref{subfig:3_188_zoom} were traced from 100 points on the Lyapunov orbit. We can observe two aspects here: first, the cylindrical shapes of these structures near the Lyapunov orbit; and second, the right branches of the invariant manifolds start inside the lunar realm, while the left ones start inside the Earth's realm.
In Fig.~\ref{subfig:3_188_full}, we show the evolution of the right branches inside the lunar realm and we note the perpendicular crossings of these structures with our surface of section. In this case, we discretized the Lyapunov orbit in 50 points to trace the invariant manifolds.

Let us now define $\Gamma$ as the intersection between the invariant manifolds and our surface of section, which can be naturally ordered by following the dynamics on $W$ and counting the crossings with $\Sigma$. We have

\be
\begin{aligned}
\Gamma^{s,u}(\alpha)=W^{s,u}(\alpha)\cap\Sigma=\bigcup_{i=1}^{\infty}\Gamma^{s,u}_i(\alpha).\\
\end{aligned}
\label{eq:manifold_crossing}
\ee

$\Gamma$ is a set of one-dimensional curves. If $\Sigma$ is always transversal to $W$, the curves are open, similar to manifolds in two-dimensional maps. Otherwise, some $\Gamma_i$ may have ellipse-like shapes as $W$ crosses $\Sigma$ in a perpendicular fashion.  Hence, the representation of invariant manifolds in phase space depends on how they intersect the surface of section.

\begin{figure*}
\centering
  \includegraphics[scale=0.75]{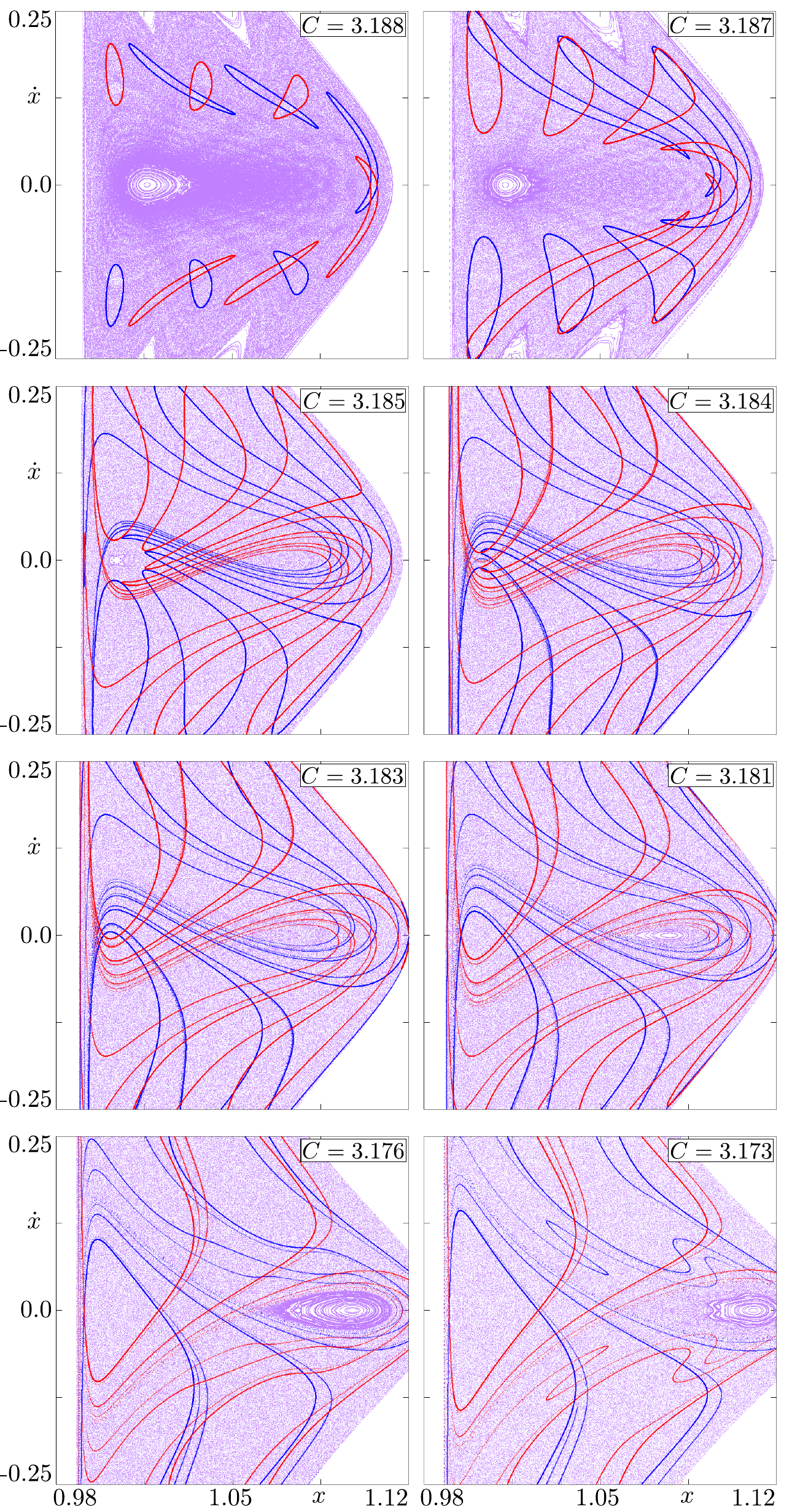}
\caption{First few components of $\Gamma^s(L)$ (blue) and $\Gamma^u(L)$ (red) in phase space. The invariant manifolds associated with the Lyapunov orbits evolve along the phase space configuration as the Jacobi constant $C$ is lowered.}
\label{fig:full_scenario}
\end{figure*}

Fig. \ref{fig:full_scenario} shows the first few $\Gamma_i(L)=W(L)\cap\Sigma$ for the same Jacobi constant values as in Fig. \ref{fig:phase_space}. The first aspect we observe is that the area enclosed by the manifolds gets bigger as we lower $C$, therefore occupying a larger region in phase space for a similar number of crossings. This is a consequence of the fact that the system is area-preserving and an element of the family of Lyapunov orbits is larger in length than the other elements with higher Jacobi constants.

The most significant result to be noted here is the fine interplay between Lyapunov orbit manifolds and phase space configuration. Initially, the manifolds intersect the surface of section far from the stability region. As we start to lower the Jacobi constant, they begin to travel across a larger area of phase space and spread towards the stability region, which gets smaller accordingly. Eventually, they cover all the stability region and the stable periodic orbit at its center bifurcates and changes stability. After the global chaos scenario, another region of stability emerges in an area of the phase space that is not yet covered by the invariant manifolds. In the end, these structures start to ripple around and invade the new stability region.

Another interesting aspect we observe from Fig. \ref{fig:full_scenario} is the apparent relationship between the spatial disposition of the invariant manifolds and the properties of the stickiness phenomenon in both mixed phase space scenarios. In Scenario I, the manifolds do not yet occupy a large portion of the phase space, which makes it possible for the stickiness to reach far into the chaotic sea. In Scenario III, on the other hand, the manifolds are spread around the new stability region and the stickiness is then confined next to it.

\begin{figure*}
\centering
\subfloat[$C = 3.187$]{\includegraphics[scale=0.7]{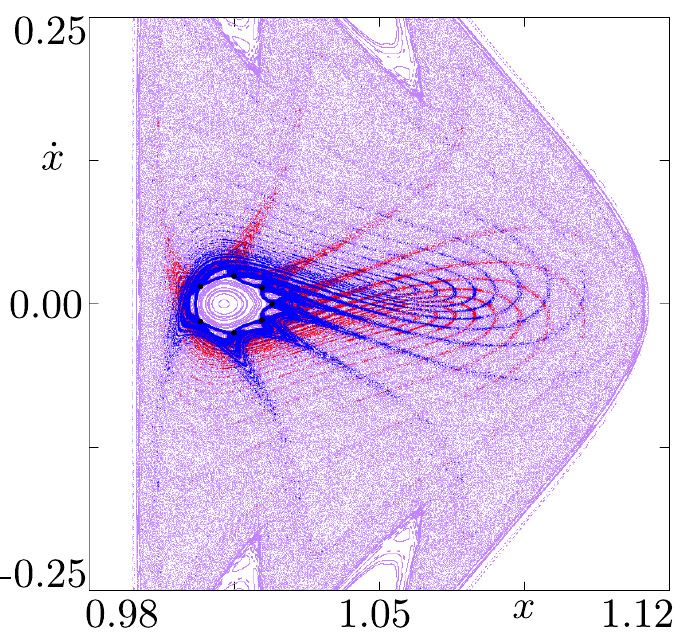}\label{subfig:first_mixed_manifolds}}
\subfloat[$C = 3.187$]{\includegraphics[scale=0.7]{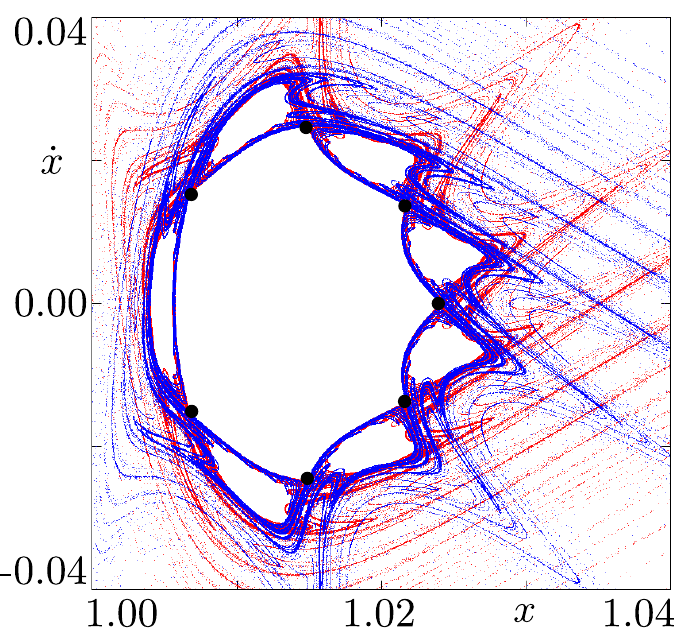}\label{subfig:first_mixed_manifolds_zoom}}
\\
\subfloat[$C = 3.176$]{\includegraphics[scale=0.7]{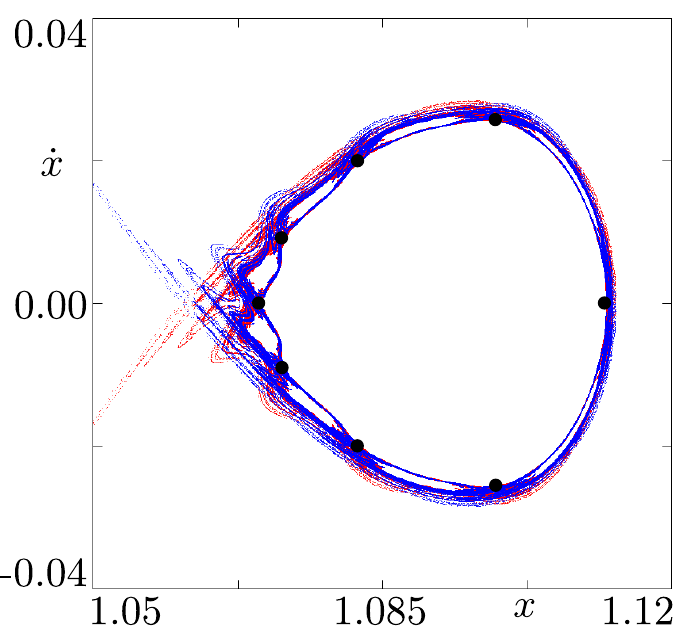}\label{subfig:second_mixed_manifolds}}
\subfloat[$C = 3.181$]{\includegraphics[scale=0.7]{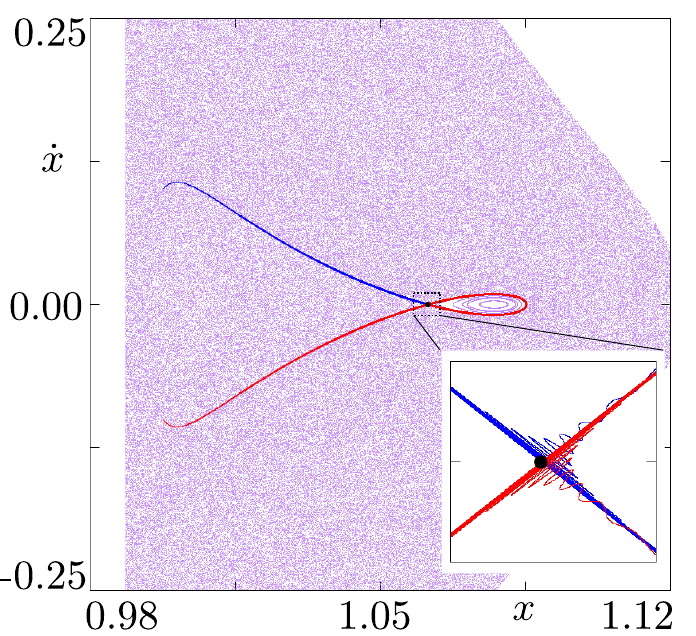}\label{subfig:saddle_manifolds}}
\caption{Stable (blue) and unstable (red) manifolds associated with the main unstable periodic orbits (black) in the mixed phase space scenarios. In Scenario I, we have $\Gamma(P^7_{I})$ in \protect\subref{subfig:first_mixed_manifolds} full size and \protect\subref{subfig:first_mixed_manifolds_zoom} zoomed-in. In Scenario III, we have \protect\subref{subfig:second_mixed_manifolds} $\Gamma(P^8_{III})$ and \protect\subref{subfig:saddle_manifolds} $\Gamma(P^1_{III})$.}
\label{fig:extra_manifolds}
\end{figure*}

As we discussed before, the stickiness effect is likely caused by invariant manifolds associated with particular unstable periodic orbits in phase space. In order to verify this assertion, we choose suitable values of $C$ for both mixed scenarios and we calculate the main unstable periodic orbit located around each regular region, which were formed from the destruction of the last KAM torus. We then trace the invariant manifolds associated with these orbits and compare them to the stickiness observed in Fig. \ref{fig:phase_space}.
For Scenario I, we choose $C=3.187$ and we calculate an unstable periodic orbit of period 7 which we call $P^7_I$. For Scenario III, $C=3.176$ and the orbit $P^8_{III}$ has period 8. 
The results are shown in Fig. \ref{fig:extra_manifolds}.

In Figs. \ref{subfig:first_mixed_manifolds} and \ref{subfig:first_mixed_manifolds_zoom}, we observe that $\Gamma(P^{7}_I)$ extend deep into the chaotic sea and closely reproduce the structure corresponding to the stickiness effect. Furthermore, Fig. \ref{subfig:second_mixed_manifolds} shows that $\Gamma(P^{8}_{III})$ are concentrated around the stability region, as we expected, also reproducing the stickiness behavior.
In Fig. \ref{subfig:saddle_manifolds}, we present the manifolds associated with the unstable periodic orbit of period 1, $P^{1}_{III}$, that is created after the second bifurcation at $C^2_{bif}$. The value of the Jacobi constant here is $C=3.181$ and we observe that $\Gamma(P^{1}_{III})$ do not have a complex geometry apart from the small oscillation near the saddle. However, it is interesting to note that a ghost effect is observed before the bifurcation with the same shape as given by these manifolds, as we can see in Fig. \ref{fig:phase_space} for $C=3.183$.

\begin{figure*}
\centering
\subfloat[$C = 3.187$]{\includegraphics[scale=0.7]{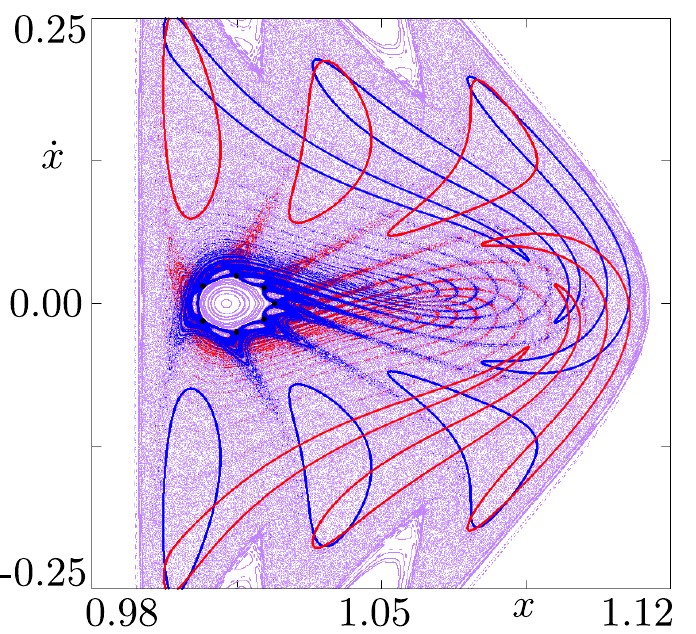}\label{subfig:all_manifolds_inner}}
\subfloat[$C = 3.176$]{\includegraphics[scale=0.7]{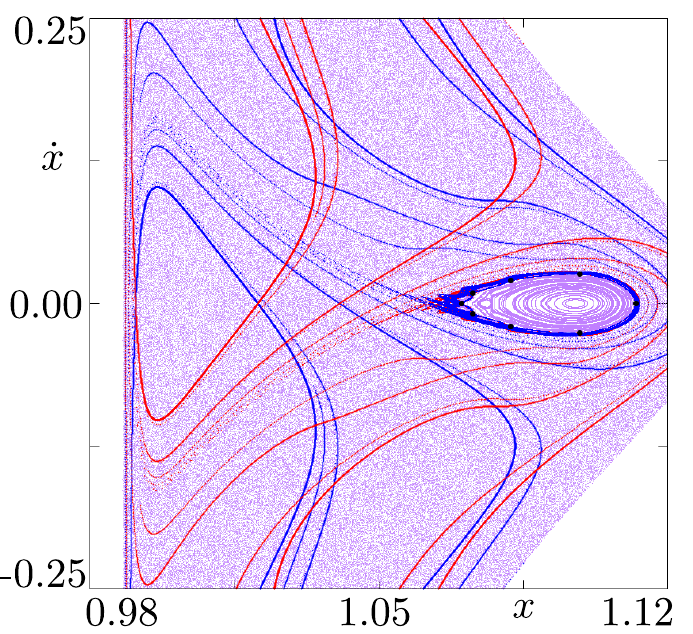}\label{subfig:all_manifolds_outer}}
\caption{Overview of the system's geometrical structures in phase space for \protect\subref{subfig:all_manifolds_inner} Scenario I and for \protect\subref{subfig:all_manifolds_outer} Scenario III. Stable manifolds are depicted in blue and unstable manifolds in red. These structures have a close relation to the phase space configuration.}
\label{fig:all_manifolds}
\end{figure*}

Finally, we depict an overview of the system in Fig. \ref{fig:all_manifolds} for the chosen Jacobi constant in each mixed phase space scenario. It is clear that each group of invariant manifolds contribute differently to the phase space configuration and that all of them are necessary for a broad description of the system.

\section{Transport analysis}
\label{sec:transport}

Another aspect regarding the phase space configuration is the presence of less dense areas in the chaotic sea. We can observe it more clearly in Fig. \ref{fig:phase_space} for $C=3.188$. If we compare it to Fig. \ref{fig:full_scenario}, we note that the less dense areas are the ones enclosed by the traced manifolds. This phenomenon comes from the fact that $W(L)$ are responsible for transporting orbits between the Moon's and Earth's realms \citep{Koon2008}. The orbits inside the first few $\Gamma^s_i(L)$ go through the Lyapunov orbit onto the Earth's vicinity faster than other areas and hence they are less populated in phase space.

\begin{figure*}
\centering
\subfloat[Forward]{\includegraphics[scale=0.7]{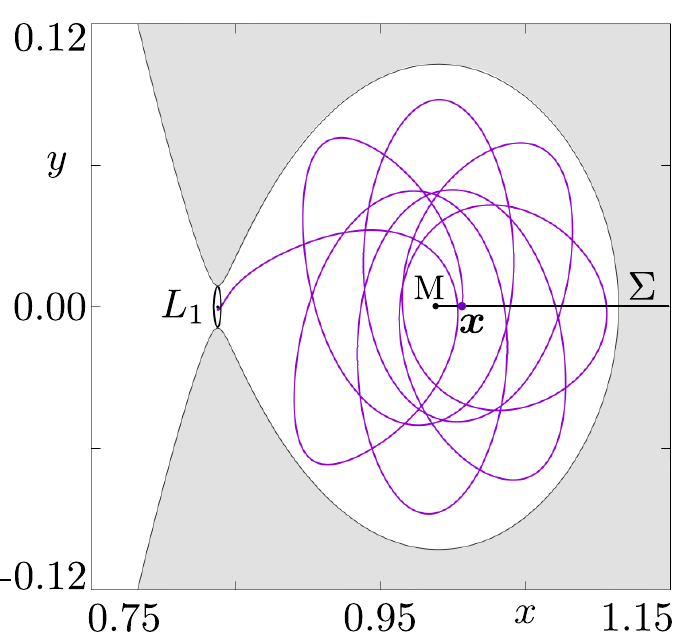}\label{subfig:orbit_fw}} \ \  \ \ \ 
\subfloat[Backward]{\includegraphics[scale=0.7]{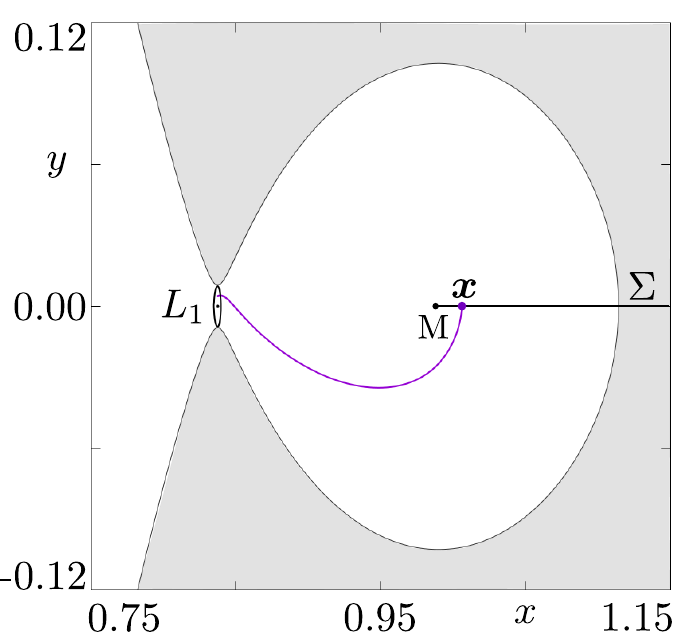}\label{subfig:orbit_bw}}
\caption{Orbit with initial condition $\boldsymbol{x}\in\Sigma$ exiting the Moon's realm for $C=3.188$. The system is integrated \protect\subref{subfig:orbit_fw} forward and \protect\subref{subfig:orbit_bw} backward on time until the trajectory enters the Earth's realm. The integration time in each case is given by \protect\subref{subfig:orbit_fw} $t_f>0$ and \protect\subref{subfig:orbit_bw} $t_b<0$. Our transit time is defined as $|t_f \times t_b|$.}
\label{fig:orbit_transit_time}
\end{figure*}

In order to dynamically quantify the geometric structures of the system, we choose orbits that begin in our surface of section and calculate how long it takes for each of them to transfer to the Earth's realm both forward $t_f$ and backward $t_b$ in time. We then define \emph{transit time} as the absolute value of the product of $t_f$ and $t_b$. An example is given in Fig.~\ref{fig:orbit_transit_time}. This is a convenient definition because our transit time highlights orbits that stay inside the lunar realm for a very long time and also for a very short time.

\begin{figure*}
\centering
\includegraphics[scale=0.75]{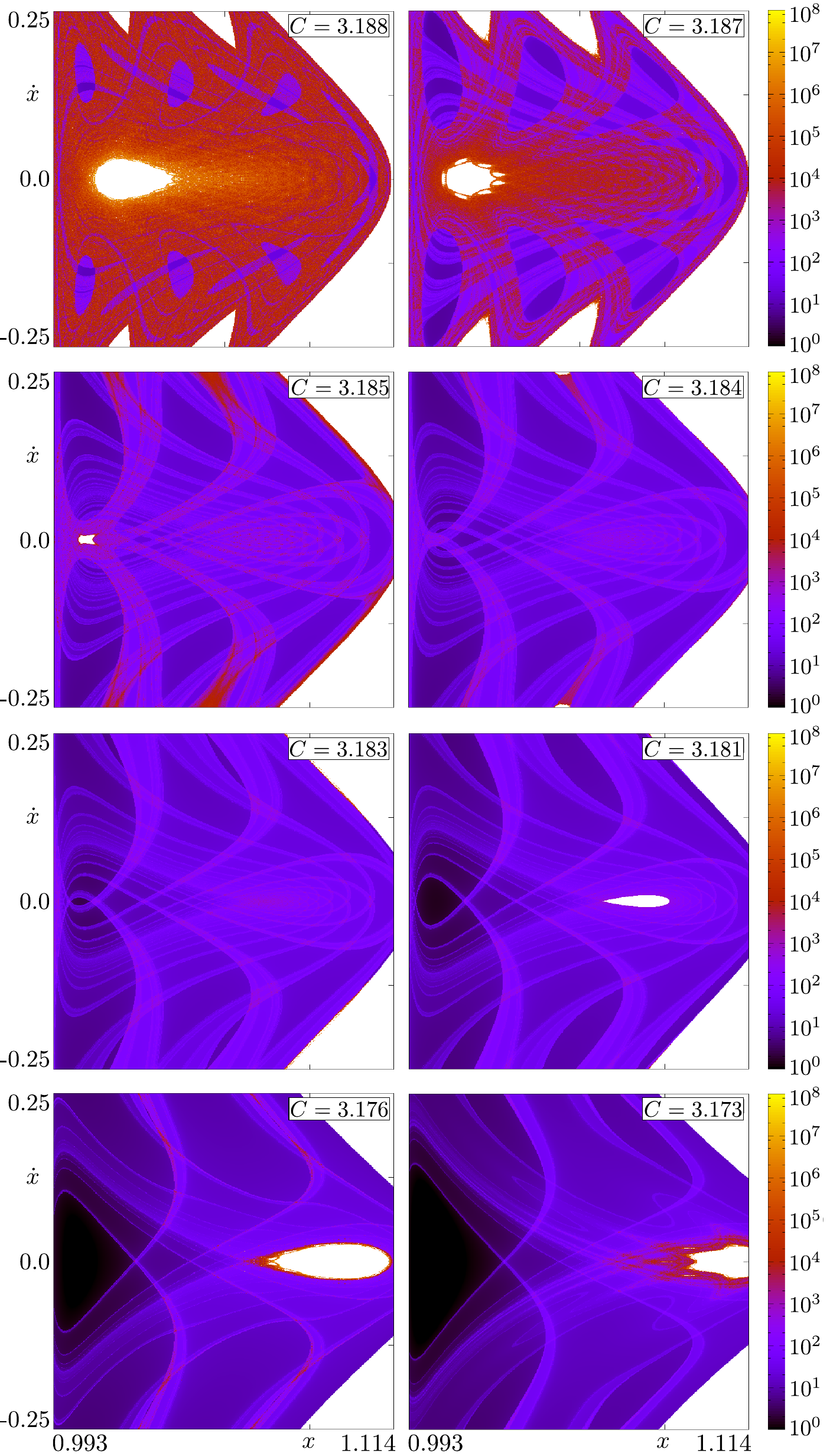}
\caption{Profile of the transit time on a logarithmic scale for different Jacobi constants. The initial conditions are chosen in the surface of section $\Sigma$.}
\label{fig:transit_time_no_collision}
\end{figure*}

Fig. \ref{fig:transit_time_no_collision} shows the transit time for a grid of $512\times1024$ initial conditions in $\Sigma$ and the same Jacobi constants of Figs. \ref{fig:phase_space} and \ref{fig:full_scenario}. The system is integrated up to $t=\pm~5\times10^{3}$ and only orbits which do eventually exit the lunar realm are considered for analysis.
We can readily observe the influence of invariant manifolds in the system's dynamics. Regions with shorter transit times correspond exactly to the interior of $\Gamma(L)$, especially inside the intersections between $\Gamma^s(L)$ and $\Gamma^u(L)$ for these are the orbits that most rapidly enter and exit the Moon's realm. In addition, regions with longer transit times correspond to the invariant manifolds associated with the main unstable periodic orbits in the mixed phase space scenarios, namely $\Gamma(P^7_I)$ and $\Gamma(P^8_{III})$.

Hence, what we observe is the coexistence of two effects. On the one hand, we have the Lyapunov orbit manifolds which are responsible for the transport between the Moon's and Earth's realms and, on the other hand, we have the manifolds associated with higher-order unstable periodic orbits which are accountable for dynamically trapping the orbits.

\begin{figure*}
\centering
\subfloat[$C = 3.188$]{\includegraphics[scale=0.7]{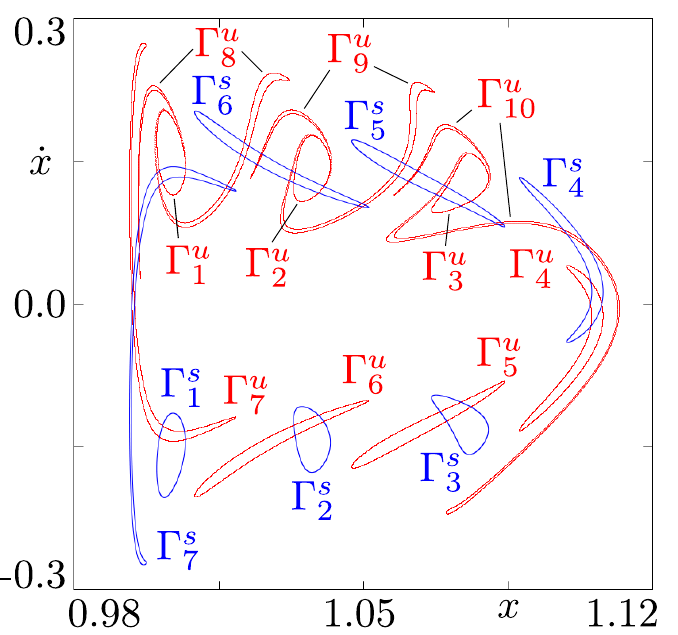}\label{subfig:break_3_188}} \ \  \ \ \ 
\subfloat[$C = 3.175$]{\includegraphics[scale=0.7]{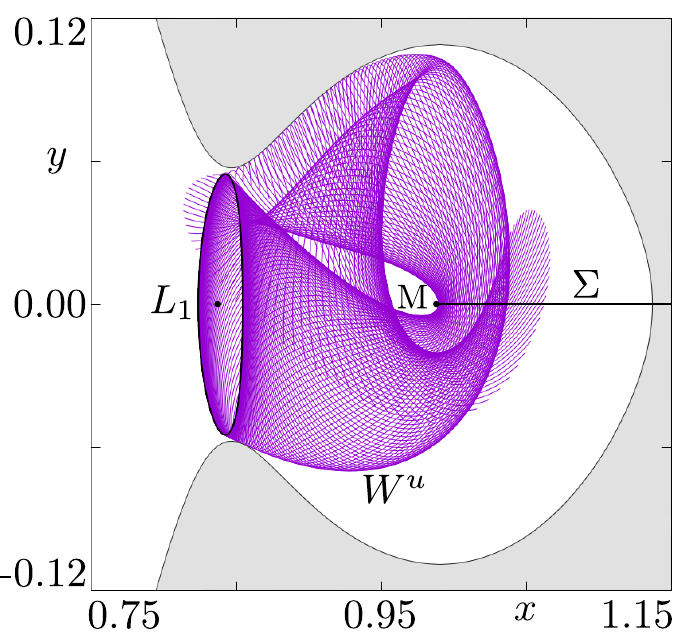}\label{subfig:break_3_175}}
\caption{Intersect and break process in the Lyapunov orbit manifolds as seen from the \protect\subref{subfig:break_3_188} phase space and the \protect\subref{subfig:break_3_175} coordinate space for different Jacobi constants. The unstable manifold eventually breaks if it intersects the stable manifold. One part of it moves to the Earth's realm while another part of it crosses $\Sigma$ again divided in two pieces.}
\label{fig:manifold_break}
\end{figure*}

All orbits in the chaotic sea, except for a set of measure zero, move from one realm to the other for a large enough integration time, which suggests that the Lyapunov orbit manifolds are dense in this area. 
The first few $\Gamma_i(L)$ are homeomorphic to circles but they eventually lose this property \citep{Gidea2007}. This phenomenon is the outcome of the intersection between two-dimensional manifolds of different stabilities. We explore this further in Fig. \ref{fig:manifold_break}.

Fig. \ref{subfig:break_3_188} shows the Lyapunov orbit manifolds in phase space for $C=3.188$. We observe that the first crossing of the unstable manifold $\Gamma^u_1$ intersects the seventh crossing of the stable manifold $\Gamma^s_7$. But, since all orbits inside $W^s$ will at some time go through the Lyapunov orbit, the intersection between  $W^s$ and  $W^u$ has the following consequence. After the seventh crossing with $\Sigma$, the orbits that compose $W^u$ are divided in three parts:
 the ones that are inside $W^s$ when the intersection occurs go through the Lyapunov orbit and on to the other realm; the ones that are exactly in the stable manifold are the homoclinic orbits and go to the Lyapunov orbit; the rest of the orbits cross the defined surface of section again $\Gamma^u_8$ although this time divided in two pieces that asymptotically approach $\Gamma^u_1$.

The described process happens indefinitely for all intersections between the unstable and stable manifolds which, by consequence, fill the chaotic sea. In Fig. \ref{subfig:break_3_175} we present the same scenario for $C=3.175$ but now in coordinate space. In this situation, both manifolds intersect each other at the first crossing and hence the unstable manifold breaks much faster. We can see a part of the manifold crossing the Lyapunov orbit whilst the other part revolves around the Moon and crosses $\Sigma$ again.

The structures that emerge from the intersect and break process are visible in Fig. \ref{fig:transit_time_no_collision}, specially for $C=3.188$. Furthermore, it is interesting to note that a somewhat similar situation occurs with the Lyapunov orbit manifolds and those associated with the higher-order unstable periodic orbits, since these structures also intersect each other. For $C=3.187$, for example, we can observe the auto-similar structure formed by the intersection between $W(L)$ and $W(P^7_{I})$.

Our final step is to examine what happens when we consider collisions with the primaries in our model. Since the structures formed by the invariant manifolds are closely related to the dynamical properties of the system, it is important for us to understand their role in this case.
In order to mimic the effects of a collision, we define a radius by hand around the Moon and stops the integration if an orbit reaches this region.
In practice, this added feature works as leaking \citep{Assis2014} for these orbits have a finite existence and therefore do not contribute to our analysis.

\begin{figure*}
\centering
\includegraphics[scale=0.75]{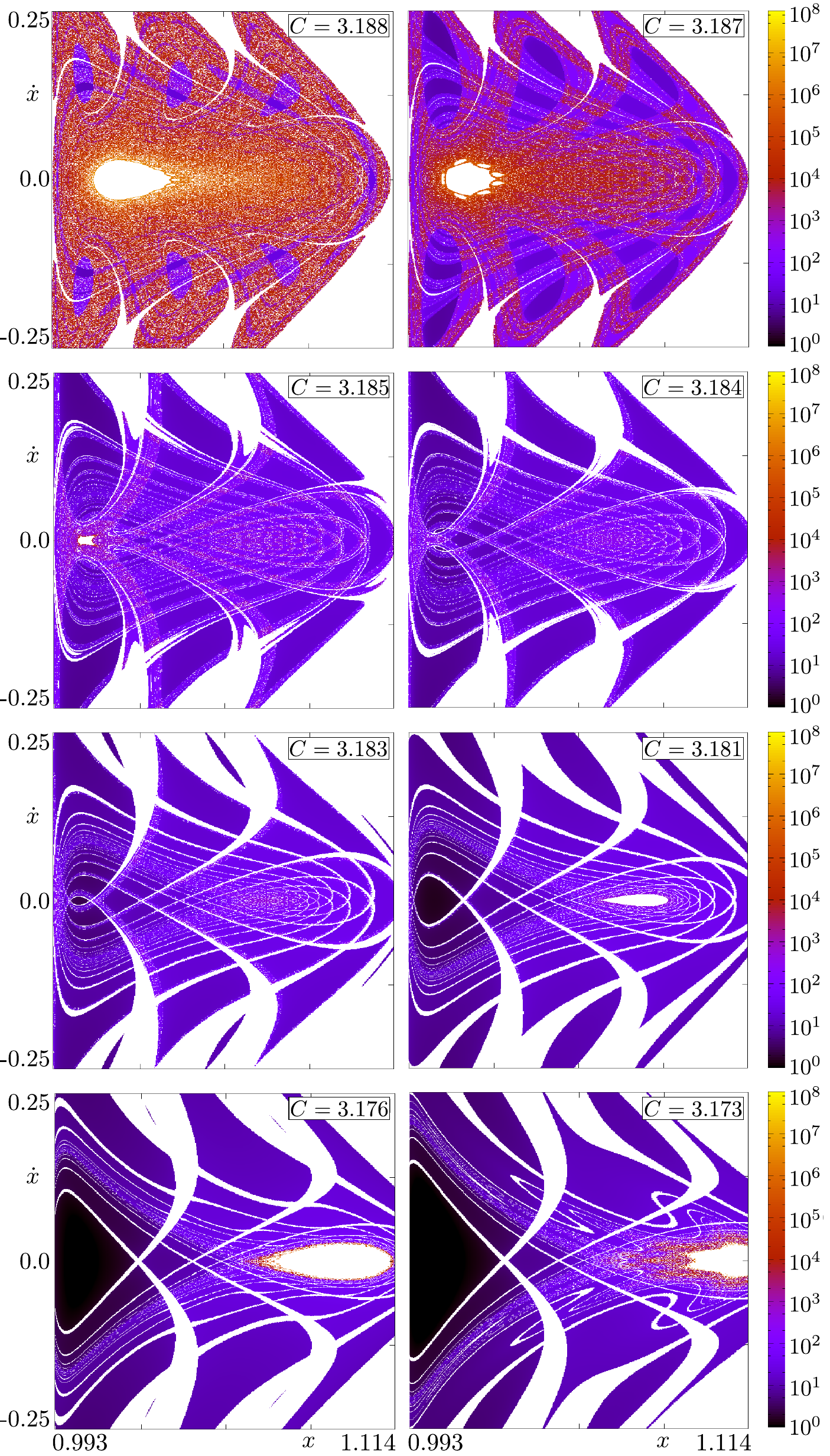}
\caption{Profile of the transit time on a logarithmic scale for different Jacobi constants, but this time discarding collisional orbits. The initial conditions are chosen in the surface of section and are the same as in Fig. \protect\ref{fig:transit_time_no_collision}.}
\label{fig:transit_time_collision}
\end{figure*}

We present the transit time profiles for this situation in Fig. \ref{fig:transit_time_collision}. The parameters chosen are the same as before and the radius of collision with the Moon is given by $r_M=4.52\times10^{-3}$. By comparison to Fig. \ref{fig:transit_time_no_collision}, we can see that the presence of a collision radius affects the dynamics of the system in two different ways. First, there is a riddled structure formed by the collisional orbits which initially covers all the analyzed space and, as we lower the Jacobi constant, it becomes more localized, mostly around the new stability region. This scheme shows a close relationship between the riddled structure and the manifolds associated with the main unstable periodic orbits in the mixed phase space scenarios.

The second effect is the appearance of collision areas which grow larger as we lower $C$, delimiting the space available to the riddled structure. Analogously, this scheme shows a close relation between collision areas and the invariant manifolds of the Lyapunov orbit. It is worth noting that if we had considered collisions in Fig.~\ref{fig:full_scenario}, for example, there would be parts of the manifolds missing and therefore their relation to the phase space configuration would be harder to visualize.

\section{Conclusions}
\label{sec:conclusions}

In this work, we showed that the planar Earth-Moon system, as modeled by the restricted three-body problem, presents three different scenarios, each one with its particular dynamical and geometrical properties.
Even though the Hill region remains topologically unchanged, the system goes from a mixed scenario with far-reaching stickiness, to the absence of stability regions, and back to a mixed scenario but now with localized stickiness, just by varying the Jacobi constant. Moreover, the transition between these scenarios are given by two different type of bifurcations, namely, the direct or inverse and the saddle-node bifurcation.

We also illustrated how some hyperbolic invariant manifolds in the system evolve along with the phase space configuration. On the one hand, we have the manifolds associated with the Lyapunov orbits, which determine the shape and size of stability regions. On the other hand, there are particular unstable periodic orbits whose invariant manifolds determine the behavior of stickiness.
These groups of manifolds are all two-dimensional surfaces, although they cross the unidimensional surface of section in different manners, hence defining geometrical structures with different properties. 

Lastly, with a reasonable definition of transit time, we were able to depict the influence of the invariant manifolds in the system's transport properties. We observed a fine interplay between the Lyapunov orbit manifolds, which are responsible for the motion between the realms, and the ones associated with the higher-order unstable periodic orbits, which temporarily trap the orbits near the stability regions.
In summary, this work provided a broad picture of the dynamics of the planar Earth-Moon system and reinforced the importance of better understanding the connection between dynamics and geometry.

\begin{acknowledgements}
VMO would like to thank Prof. Dr. J. D. Mireles James for his notes on Celestial Mechanics.\footnote{Available at \url{http://cosweb1.fau.edu/~jmirelesjames/notes.html}.} This study was financed in part by the Coordena\c c\~ao de Aperfeiçoamento de Pessoal de Nível Superior - Brasil (CAPES) - Finance Code 001 and the São Paulo Research Foundation (FAPESP, Brazil), under Grant No.  2018/03211-6.
\end{acknowledgements}

\bibliographystyle{myspbasic}      

\bibliography{manuscript_citation}   

\end{document}